\documentclass[showkeys,reprint,amsmath,amssymb,aps]{revtex4-2}

\usepackage{graphicx}
\usepackage{orcidlink}
\usepackage{subcaption}
\usepackage{physics}
\usepackage{float}

\begin{document}

% \preprint{APS/123-QED}

\title{Coupling nitrogen-vacancy centre spins in diamond to a grape dimer}

\author{\orcidlink{0009-0007-7498-0741}Ali Fawaz}
\email{ali.fawaz@hdr.mq.edu.au}
\author{\orcidlink{0000-0002-9081-5750}Sarath Raman Nair}%
 \email{sarath.raman-nair@mq.edu.au}
 \author{\orcidlink{0000-0001-9850-4992}Thomas Volz}%
 
\affiliation{%
 School of Mathematical and Physical Sciences, Macquarie University, North Ryde, 2109, NSW, Australia
}%
\affiliation{Centre of Excellence in Engineered Quantum Systems, North Ryde, 2109, NSW, Australia}


\begin{abstract}
Two grapes irradiated inside a microwave (MW) oven typically produce a series of sparks and can ignite a violent plasma.
The underlying cause of the plasma has been attributed to the formation of morphological-dependent resonances (MDRs) in the aqueous dielectric dimers that lead to the generation of a strong evanescent MW hotspot between them. 
Previous experiments have focused on the electric-field component of the field as the driving force behind the plasma ignition.
Here we couple an ensemble of nitrogen-vacancy (NV) spins in nanodiamonds (NDs) to the magnetic-field component of the dimer MW field.
We demonstrate the efficient coupling of the NV spins to the MW magnetic-field hotspot formed between the grape dimers using Optically Detected Magnetic Resonance (ODMR).
The ODMR measurements are performed by coupling NV spins in NDs to the evanescent MW fields of a copper wire. 
When placing a pair of grapes around the NDs and matching the ND position with the expected magnetic-field hotspot, we see an enhancement in the ODMR contrast by more than a factor of two compared to the measurements without grapes.
Using finite-element modelling, we attribute our experimental observation of the field enhancement to the MW hotspot formation between the grape dimers.
The present study not only validates previous work on understanding grape-dimer resonator geometries, but it also opens up a new avenue for exploring novel MW resonator designs for quantum technologies.
\end{abstract}

%Use showkeys class option if keyword
\maketitle

%\tableofcontents

\section{Introduction}

Grapes are one of the most popular fruits with a lot of health benefits
\cite{yang_grape_2013}.
Since the first observation of sparks in the gap between two spatially-separated grape pieces inside a microwave (MW) oven in 1994 \cite{michaud_fun_1994}, grapes became a central character in an interesting physics problem. 
Khattak et al. \cite{khattak_linking_2019} provided a physical explanation for the rapid succession of sparks phenomenon by focusing on the electric field distribution around the grapes.
They show that the grape dimers (or hydrogel water beads with implanted ions) act like a MW resonator where the fields are stored inside the grapes due to morphological-dependent resonances (MDRs).
MDRs in grapes are caused by their curvature \cite{affolter_electromagnetic_1973} and high-electric permittivity (e.g. Re$(\epsilon_{r})\approx 80$ at 2.5 GHz \cite{meissner_complex_2004}). The observed sparks are due to the plasma formation from metallic ions in the grapes \cite{lin_electromagnetic_2021}. 
Motivated by this work, one might wonder whether such grape dimer arrangements could enable technical applications where strong MW field enhancement is needed.

\begin{figure*}
\centering
\includegraphics[width=18cm]{"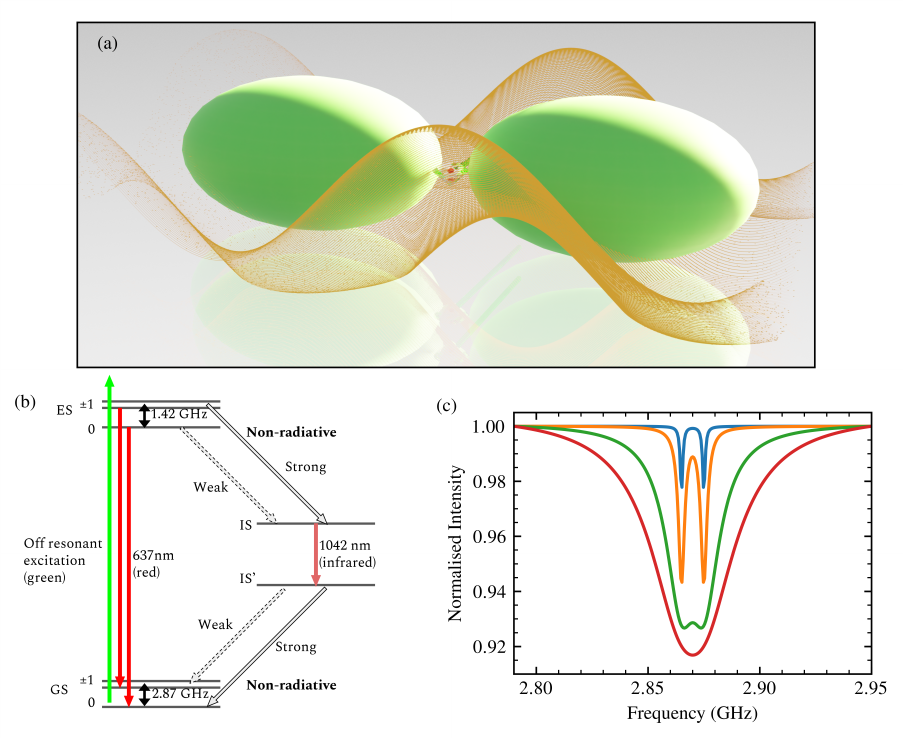"}
\caption{(a) Artistic representation of the central ideal of the present paper: A ND (red) containing an ensemble of NV spins is placed in the gap of a grape dimer (green). The NV spins exchange MW photons with the confined field mode between the grapes, indicated by the orange field lines. (b) NV$^{-}$ internal level structure showing the triplet (ES, GS) and intermediate singlet state (IS, IS') manifolds. The zero-field ground-state splitting (between $m_{s}=0$ and $m_{s}=\pm 1$) is approximately 2.87~GHz. Grey-shaded arrows represent inter-system crossing (ISC) transitions ES to IS and IS' to GS. (c) Simulated ODMR profile of NV$^{-}$ for different magnetic-field coupling strengths with blue representing weak coupling, orange and green representing intermediate coupling and red representing strong coupling; refer to Appendix A for calculations.}
\label{fig:1}
\end{figure*}

Microwave resonators that confine MW fields to a finite volume are employed in satellite technology, maser technology, microwave photon detection \cite{lescanne_irreversible_2020}, dark matter axion search \cite{bradley_microwave_2003}, quantum technologies \cite{probst_three-dimensional_2014} \cite{eisenach_cavity-enhanced_2021} \cite{young_cavity_2009}, etc.
In the context of quantum technologies, MW resonators are used to coherently drive embedded quantum systems, such as e.g. Rydberg atoms, solid-state spins or superconducting qubits.
In the specific case of spin qubits, it is the magnetic field of the MW resonator that couples to the magnetic moment associated with the spin. 
Here we use the enhanced magnetic field in grape dimers to efficiently drive nitrogen-vacancy (NV) centre spins in nanodiamonds (NDs). An artistic representation is shown in Fig. \ref{fig:1}(a).

The NV spins are solid-state quantum systems whose intrinsic coherence times can reach millisecond timescale, even at room temperature \cite{stanwix_coherence_2010}\cite{herbschleb_ultra-long_2019}.
Their preparation and readout can be done by detecting spontaneously emitted fluorescence upon illuminating the diamond with non-resonant green light \cite{gruber_scanning_1997}.
The possibility to optically probe the NV spins at room-temperature enable us to easily incorporate the quantum systems into the grape cavity. 
Furthermore, we consider NV spins in diamond as the right choice for this demonstration as they have proven particularly useful for sensing of magnetic fields, pressure and temperature \cite{doherty_electronic_2014} \cite{neumann_high-precision_2013}.

This article is organised as follows.
In section (2), we discuss the experimental setup for studying the coupling of NV spins in NDs to the magnetic mode of the grape resonator and carrying out ODMR measurements to demonstrate the effect of the grape resonator.
Then in section (3), we investigate the hotspot formation in the grape resonator using finite element simulation.
Finally, in section (4), we conclude the presented study.

\section{Experimental coupling of NV spins in NDs to grape dimer resonators}\label{sec2}

We use real grapes to experimentally demonstrate improved MW field interactions with NV spins. This section presents a brief background on NV spin, the experimental technique used, and the measurements performed with grape dimers.

\subsection{NV spins}

NV spins are one of the many possible point defects that can exist in diamond. In its negative charge state NV spins which we denote as NV$^{-}$ spins and they are interesting for room-temperature quantum applications. NV$^{-}$ is a spin - 1 system, which has triplet ground and excited states. The room-temperature energy structure of NV$^{-}$ is shown in Fig. \ref{fig:1}(b). The ground and excited states of NV$^{-}$ are denoted as GS and ES, respectively. Being triplet states, ground and excited states each consist of three states with magnetic spin quantum numbers $m_{s}=0,\pm1$. 

\begin{figure*}
\centering
\includegraphics[width=18cm]{"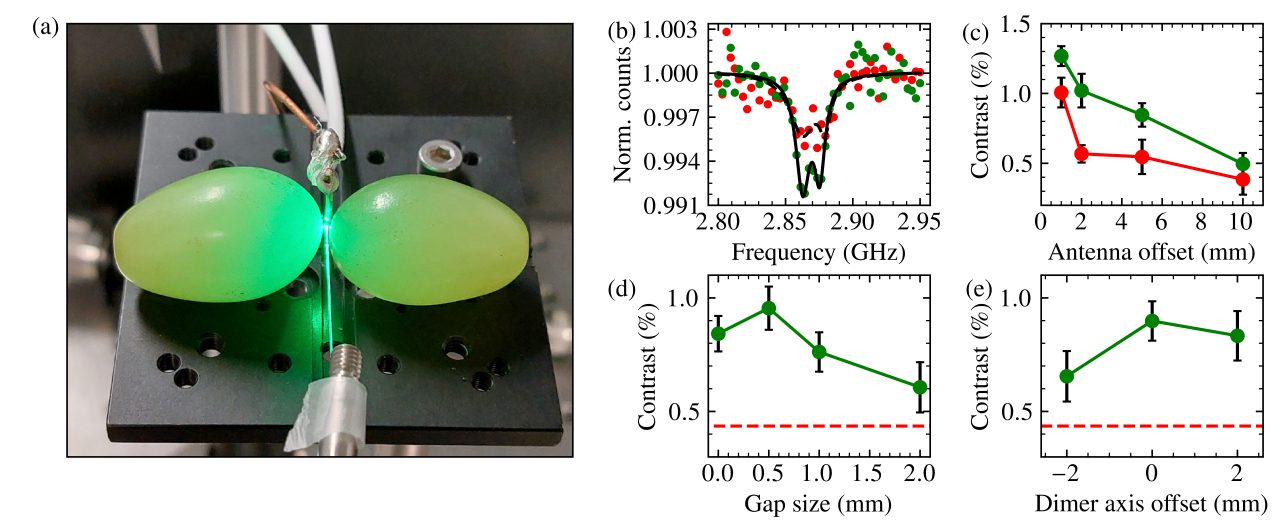"}
\caption{(a) Photo of the experimental setup to couple MWs to NVs using grape dimers. A stripped optical fibre with NV spins, cantilevered from a rod, lies between two grapes. The grapes were positioned on a platform with a vertical straight copper wire, equidistant from each grape. (b) Typical ODMR measured with (green) and without (red) grape dimers showed improved ODMR contrast. Double-Lorentzian fits (black curves) were used to extract the contrast. The dashed black curve is the fit for the case without grapes and the solid black curve is the fit for the case with grapes. (c) ODMR contrast plotted as a function of the distance from the antenna to the centre of the gap, with a fixed gap size of 1~mm. (d) Contrast plotted as a function of gap size, with the distance of the wire antenna fixed to 5~mm from the gap. (e) ODMR contrast plotted as a function of the distance from the dimer axis near the gap with the grapes and antenna fixed. The antenna was positioned 5~mm from the gap, and the gap size was fixed to 1~mm.}
\label{fig:3}
\end{figure*}

Using an off-resonant optical driving with a green laser (e.g. 532~nm), NV$^{-}$ can be optically polarised to the $m_{s}=0$ ground state. The green laser excites NV$^{-}$ spins from GS to ES conserving the $m_{s}$  (0 to 0 and $\pm$1 to $\pm$1). The NV$^{-}$ spins relax radiatively from ES to GS by emitting broad red light spectrum from $\sim$600~nm -- 800~nm with a zero phonon line transition around $\sim637$~nm. NV$^{-}$ spins also undergo a non-radiative transition known as inter-system crossing (ISC) to the intermediate singlet state manifold (labelled IS). The ISC transitions from ES ($m_{s}=0$) $\rightarrow$ IS and ES ($m_{s}=\pm 1$)$\rightarrow$ IS occur at different rates, with the $m_{s}=\pm1$ transition occurring more frequently \cite{gupta_efficient_2016}. The spin is short-lived in IS compared to other transitions \cite{ulbricht_excited-state_2018} and relaxes to IS' by emitting infrared light ($\sim$1042~nm). Non-radiative transitions to the ground state IS' $\rightarrow$ GS $(m_{s}=0)$ and IS'$\rightarrow$ GS $(m_{s}=\pm1)$ also occur at different rates, with the $m_{s}=0$ transition being more favourable \cite{gupta_efficient_2016}. Consequently, due to the varying rates of the ISC levels from the $m_{s}=0,\pm1$ levels, the spin eventually becomes polarised in the ground state GS ($m_{s}=0$) and repeatedly cycles through excitation and de-excitation. 

The states $m_{s}=\pm1$ are typically degenerate in the ground state. The degeneracy can be lifted by applying a magnetic field along the NV axis. The degeneracy can also be inherently lifted due to lattice strain \cite{doherty_nitrogen-vacancy_2013}. The zero-field splitting transition from $m_{s}=0 \rightarrow m_{s}=\pm1$ corresponds to a MW transition with frequency $2.87$~GHz. With optical pumping polarising the spin in $m_{s}=0$, a MW field (near 2.87~GHz) creates a population change in the ground triplet states. Due to different rates of ISC transitions from ES ($m_{s}=0$) and ES ($m_{s}=\pm1$), the MW field indirectly induces a reduction in the red light emitted from the NV spin. This process is referred to as optically detected magnetic resonance (ODMR), where magnetic dipole transitions in the MW domain are examined through the observed reduction of optical photons; see Fig. \ref{fig:1}(c) for a typical ODMR profile with non-degenerate $m_{s}=\pm1$ levels. The contrast of the ODMR plot represents the percentage reduction in red light intensity in the ODMR profile, depending on the strength of the MW field (see Appendix A for numerical modelling of ODMR and MW field strength). 

\subsection{Experimental setup}

We experimentally measure the ODMR of NV$^{-}$ spins in NDs in the presence of grape dimers and without grapes. In order to have a small probe volume and efficient optical access, the NDs with the NV centres are attached to the tip of a multimode optical fibre. A 1~mm diameter vertical copper wire served as a MW antenna, driven by a MW generator (Agilent Technologies E8257D) and a 16W-amplifier (Minicircuits ZHL-16W-43-S+). The antenna was placed vertically such that the radiated component of the MW field points in the direction of the dimer axis. Here the dimer axis refers to the line passing through the major axis of each grape. As shown in Fig. \ref{fig:3}(a), the fibre tip is positioned in between the gap between the two grapes. To a good approximation, the grapes have an ellipsoidal shape and were chosen carefully to have approximately identical dimensions, with a long axis of about 27~mm and a short axis of about 17~mm. The size of the grapes used were selected based on numerical simulations of grape dimers (see Appendix B for detailed materials and methods).

\subsection{Enhanced ODMR measurements with a grape dimer resonator}

ODMR measurements performed with and without grape dimers showed a clear improvement in MW coupling strength, as is evident from Fig. \ref{fig:3}. Fig. \ref{fig:3}(b) shows two measured ODMR profiles for the case where the grapes are as close as possible to the NV nanodiamonds (green) and the case where there are no grapes present (red). The improvement in MW coupling is inferred from the increase in contrast. The contrast was extracted from the bare ODMR measurements using a double-Lorentzian fit taking into account the splitting between the $m_{s}=\pm1$ states. Lorentzian fits are known to fit NV ODMR profiles closely \cite{fujiwara_observation_2018}. Here we attribute the increase in contrast due to increases in the magnetic field strength in the gap. We note that other factors, such as laser power and temperature, could also affect the ODMR contrast. However, the laser power was stable such that the overall PL intensity counts remained stable within $\approx10\%$ throughout the experiments. As for thermal effects, since the grapes heat up due to the absorption of MW radiation, the environmental temperature exhibited by the NV spins is expected to be greatest when grapes are present. With increasing temperature, it has been shown \cite{liu_coherent_2019} that ODMR contrast for NVs in NDs decreases. In addition, the effect of ionisation causing a change in the charge state from NV$^{-}$ to NV$^{0}$ was also considered. As temperature increases, the ionisation rate may increase and can cause changes in the contrast. Through numerical modelling, the ionisation rate was found to have minimal effects on ODMR contrast for relatively small temperature changes. We, therefore, exclude temperature and laser power as potential causes for the increase in ODMR contrast, and instead attribute it to changes in the local magnetic field. 

Several ODMR measurements were performed to examine the dependence of different parameters and to reveal information on the MW hotspot. With the gap size fixed to 1~mm, the grapes and fibre tip were translated at different distances from the straight copper wire antenna. The results shown in Fig. \ref{fig:3}(c) indicate a reduction in contrast with increasing distance from the antenna. This is expected since the field strength decays away from the antenna. With the grape dimer present, an enhanced contrast was observed in all cases, indicating the presence of the grapes as the main cause for the increased contrast and not some favourable geometric arrangement. Due to additional heating when the grapes are present, the optical fibre may expand in length and move closer to the copper wire, increasing the contrast. However, this effect is negligible since each ODMR contrast measured with grapes is greater than the measured contrast without grapes within distances of 2mm of each point. Therefore, we also eliminate thermal expansion as the cause of the increase in contrast. Furthermore, any additional fluorescence from the optical fibre itself is expected not to influence the measured ODMR contrast.

By fixing the antenna at a distance of 5~mm orthogonal to the dimer axis and varying the gap size, we were able to experimentally determine an optimum gap size of about 0.5~mm for maximum contrast improvement (see Fig. \ref{fig:3}(d)). 
Finally, with the gap size fixed at 1~mm and the copper antenna positioned 5~mm from the grapes, the fibre tip was translated 2~mm in both directions orthogonal to the dimer axis. The plot in Fig. \ref{fig:3}(e) indicated the contrast was the largest along the dimer axis.

\section{Finite element simulation of magnetic field hotspot formation in grape dimer resonator}\label{sec3}

\begin{figure*}
\centering
\includegraphics[width=18cm]{"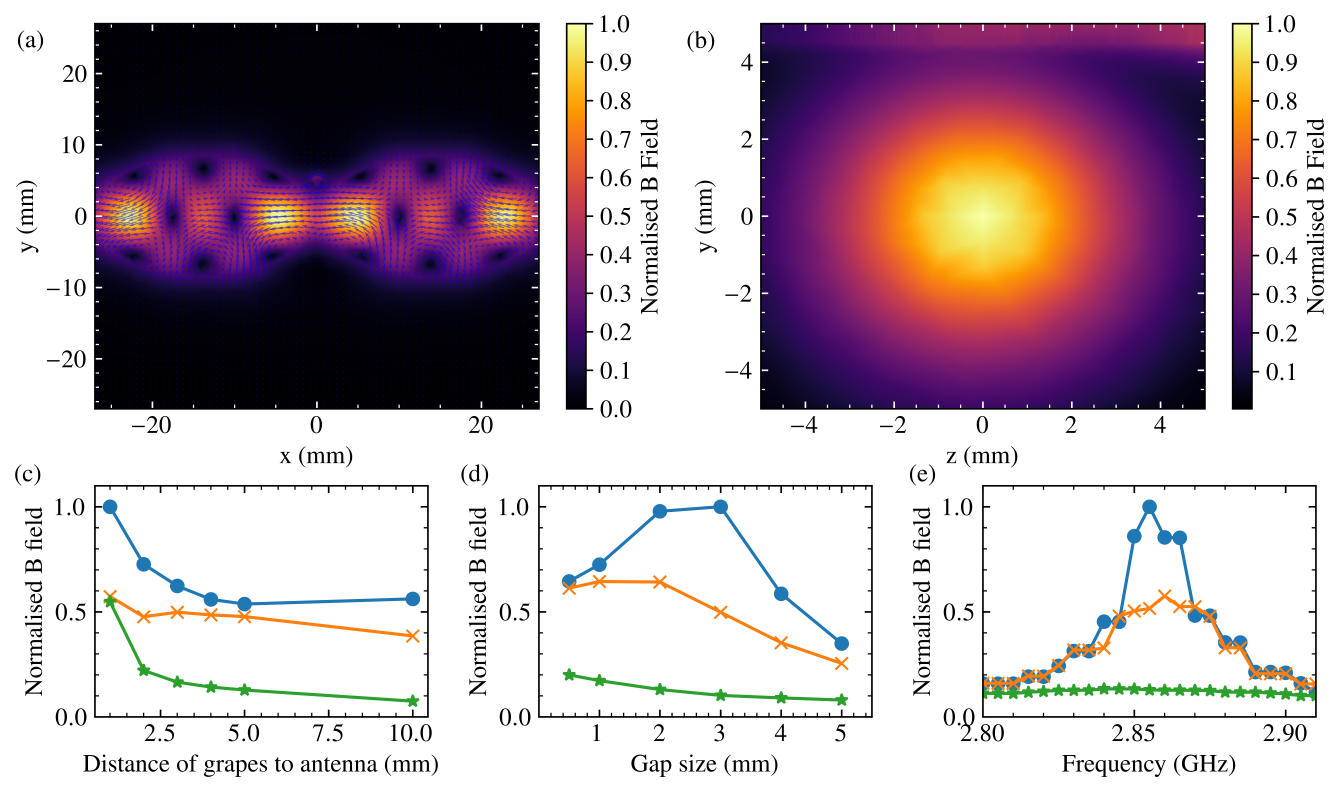"}
\caption{Results of finite element (FE) simulations for an ellipsoidal dimer with dimensions of 2.7~cm (major) and 1.7~cm (minor) at 2.87 GHz. (a) The contour of the magnetic field at a cross-section through the grapes at z=0. The vector field (blue) is also shown. (b) The contour of the magnetic field at a cross-section through the gap at x=0 clearly shows the formation of a hotspot. The magnetic field is observed to increase at the top of the plot since the antenna is located near this side. (c) The magnetic field at the centre of the hotspot is plotted for different antenna distances from the dimer axis. The different colours correspond to no absorption - blue ($\epsilon_{r}=79.21$), weak absorption - orange ($\epsilon_{r} = 79.21+1i$), and strong absorption - green ($\epsilon_{r}=79.21+5i$). The gap size was fixed to 1~mm. (d) The magnetic field at the centre of the hotspot as a function of gap size for different absorption strengths. The antenna was fixed to a distance of 5~mm from the dimer axis. (e) The magnetic field at the centre of the hotspot as a function of MW frequency for different absorption strengths. The gap size was fixed to 1~mm, and the antenna was placed at 5~mm from the dimer axis. }
\label{fig:6}
\end{figure*}

In order to verify the magnetic field hotspot between grape dimers and confirm that the increase in ODMR contrast is due to the change in the magnetic field, finite element simulations were performed. Related work \cite{khattak_linking_2019} investigated the MW hotspot formation in grapes and only focused on the effects of the electric field in the hotspot, which caused the plasma formation. 
Because the electric field is known to be large in the gap between grape dimers due to observed plasma formation, it is expected that the magnetic field should also form similar resonant behaviour in the gap. Furthermore, we simulate the MW source more realistically by defining a straight copper wire antenna (as shown in the experimental setup) instead of an ideal plane wave.

Two grapes were modelled in COMSOL 6.0. Fig \ref{fig:6} shows the FE results for simulated grape dimers. In the simulations, the grapes were modelled as ellipsoids and given the properties of water at MW frequencies ($\epsilon_{r}=79.21$) as assumed by Khattak et al. \cite{khattak_linking_2019}. The dimensions of the ellipsoids was taken as 27~mm (major axis length) and 17~mm (minor axis length). These dimensions were determined by measuring the major axis length of typical white seedless grapes purchased from a local supermarket. These grapes varied in size; however, a vast amount of the available grapes were approximately 27~mm long with varying widths. With approximate calculations from Mie theory, the corresponding minor axis dimensions of the ellipsoids were calculated to achieve resonance at roughly 2.87 GHz (see Appendix C). To deliver MW radiation to the ellipsoids, a dipole antenna with 1~mm diameter, equidistant from the dimers, was modelled to simulate a more realistic radiation source rather than just applying simple plane waves. This modelled the delivery of MW radiation more closely to the experimental setup.

The ellipsoid dimers and the antenna were embedded in an air sphere of radius 80~mm. The air sphere was surrounded by a spherical shell of thickness 53.5~mm, which acts as a perfectly matched layer (PML). PMLs are designed to absorb radiation to prevent artificial reflections due to the finite size of the model. In this case, the PML is designed as half the maximum wavelength ($\lambda\approx107$~mm) used in the simulation. Maxwell’s equations in 3D were then solved using a frequency domain analysis.

Figures \ref{fig:6}(a) and (b) show the typical magnetic field resonance patterns at 2.87~GHz for in-plane and out-of-plane cross sections, respectively. Several maxima in the interior of the ellipsoids are observed, which is suggestive of a higher-order mode resonance. The internal field in the dimers was found to exhibit larger field strengths compared to the external field in the gap. Ideally, one can attempt to couple to these interior fields; however, the direction of the magnetic field in the interior may be difficult to control and, therefore, to couple to NVs. Based on Mie theory calculations (see Appendix C) on a dielectric sphere with a cross-sectional circumference similar to the ellipsoids, the pattern resembles a spherical third-order magnetic field resonance mode. Since MDRs occur due to the total internal reflection of waves along the boundary, the resonance condition is typically proportional to the circular circumference of a sphere in Mie theory; a similar condition for ellipsoids is expected. 

A vector field distribution from FE simulations is also shown in Fig. \ref{fig:6}(a), showing the internal and external vector magnetic fields. The direction of the magnetic field in the gap plays an important role when coupling to solid-state spins. Angular deviations of the magnetic field will reduce the amount of coupling achieved. In the simulated ellipsoidal dimers, the magnetic field in the gap lies conveniently along the dimer axis. 

Simulations were performed where the antenna was translated at various distances orthogonal to the dimer axis while remaining equidistant from each grape. In Fig. \ref{fig:6}(c), the magnetic field at the centre of the hotspot was plotted as a function of the distance to the antenna for different absorption strengths. As absorption effects increased, the magnetic field in the hotspot was weaker. The FWHM of the hotspot was estimated from these simulations to be 6.2~mm, indicating some flexibility for placing the NV diamonds for achieving enhanced coupling.
As the gap size was varied, the magnetic field reached a maximum, as plotted in Fig. \ref{fig:6}(d). The magnetic field in the gap can be optimised by tuning the gap size, in agreement with the experimental findings on gap size reported earlier. 
We also checked the influence of the absorption on the microwave resonances and the strength of the magnetic field hotspot. The results are displayed in Fig. \ref{fig:6}(e). In these simulations, the gap size was fixed to 1~mm, and the dipole antenna was placed 5~mm from the dimer axis. Different absorptive levels of water were taken into account by assuming a complex electric permittivity. Since the exact absorption effects in the grapes are unknown, the complex relative electric permittivities considered were Im($\epsilon_{r}$)$=$ $0i$, $1i$, $5i$. The peak magnetic field was found to occur at $2.855$~GHz (near 2.87 GHz) with a full width at half maximum (FWHM) of approximately 25 MHz (no absorption). With increasing absorption effects, more losses are expected in the media leading to a broader resonance and a smaller peak value.

Comparison between the simulation results and the experimental findings gives a consistent picture: The formation of a magnetic-field hotspot between the grapes in our FE simulations nicely explains the measured increases in ODMR contrast as a function of gap size and spatial location. 

\section*{Conclusions}\label{sec4}

In the presence of grape dimers, NV spins were driven resonantly with MW fields and enhanced ODMR contrast was demonstrated. Our NV-ODMR technique directly detects the magnetic-field hotspot in the grape dimer gap. The results presented here further support and complement the earlier findings by Khattak et al. \cite{khattak_linking_2019} that provided an explanation for the long-standing mystery of plasma ignition by grape dimers in a microwave oven. Beyond this aspect, our work lays the foundation for the exploration of novel microwave cavity designs based on dielectric multi-spherical/ ellipsoidal geometries in the context of spin-based quantum technologies.

\begin{acknowledgments}
This research was supported by the ARC Centre of Excellence on Engineered Quantum Systems (CE170100009). A.F. acknowledges funding through a domestic MQ Research Training Program (RTP) scholarship stipend. The authors would like to thank L.~Cronin and J.~Geordy for valuable advice on the experimental aspects of this project, and M.~Schmidt and M.~Steel for providing access to the COMSOL software. We thank Commonwealth Science and Industrial Research Organisation (CSIRO - Lindfield New South Wales) for technical support. 
\end{acknowledgments}

\section*{Authors' Contributions}
Author contribution: T.V. conceived the central idea of the project; T.V. and S.R. designed the project. S.R. and A.F. designed and built the experimental setup; A.F. and S.R. performed the experiments; A.F. and S.R. analysed data; A.F. and S.R. performed theoretical modelling; A.F. conducted finite element modelling; All authors interpreted the results and wrote the manuscript. S.R. and T.V. supervised the project.

\section*{Appendices}
\appendix

\renewcommand\thefigure{\thesection.\arabic{figure}}  

\section{ODMR Simulations of NV Spins}\label{secA1}
\setcounter{equation}{0}
\setcounter{figure}{0}    
To simulate the NV spin response to MW fields (specifically, the magnetic field component), theoretical equations were derived using the quantum master equation approach. At room temperature, NW spins contain ground and excited spin-triplet states of NV$^{-}$, two intermediate states of  NV$^{-}$ and also two NV$^{0}$ states. Since the spin lifetime of ${}^{1}A_{1}$ is much smaller than the lifetime of ${}^{1}E$ (see main text), the singlet states were effectively assumed as one state (${}^{1}E$). The inclusion of NV$^{0}$ states accounts for ionisation effects responsible for charge state switching. In the NV$^{0}$ state, it can be optically pumped back to the ground triplet state of NV$^{-}$. The NV energy scheme consisting of nine levels is shown in Figure \ref{f:2_1}. Simpler models could be used with fewer levels in the case of degenerate triplet states and if ionisation effects are neglected. The essential states to model ODMR are the ground, intermediate and excited states of NV$^{-}$. However, since transition rates between NV$^{0}$ and NV$^{-}$ are known, and readily available from various sources, NV$^{0}$ levels were included in the modelling.

The essential dynamics to capture in the equations relates to the NV spin-triplet ground state, which interacts with the MW field. Magnetic dipole interactions $(\Delta m_{s}=1)$ in the ground state can be modelled as two-level quantum emitters \cite{wu_superradiant_2022}. Since the microwave field primarily induces spin transitions $m_{s}=0\leftrightarrow1$ and $m_{s}=0\leftrightarrow-1$, each transition was treated as a two-level quantum emitter. Furthermore, nuclear and spin-spin interactions were neglected as such interactions can be effectively treated as a source of dephasing, specified in the form of a Linbladian dephasing operator for simplicity. Since exact information about the grape cavity is unknown, the Hamiltonian was modelled as an ensemble of two-level quantum emitters interacting with a classical AC magnetic field. The driving strength $\Omega$ between the degenerate $m_{s}=\pm1$ levels and the $m_{s}=0$ were assumed to be identical for simplicity. The Hamiltonian (using the rotating wave approximation) for a single NV spin is given in Equation \ref{eq:2_1}. Note that since spin-spin interactions are neglected, the Hamiltonian for each spin is the same, so we can analyse one spin to determine the effects of magnetic field strength on the NV-ODMR response.

\begin{equation}
\hat{H}/\hbar=\Delta_{12}\hat{\sigma}_{22}+\Delta_{13}\hat{\sigma}_{33}+\Omega(\hat{\sigma}_{21}+\hat{\sigma}_{12}+\hat{\sigma}_{13}+\hat{\sigma}_{31})
\label{eq:2_1}
\end{equation}

In this Hamiltonian, $\Delta_{12}=\omega_{12}-\omega_{mw}$ and $\Delta_{13}=\omega_{13}-\omega_{mw}$ are the detuning frequencies of the corresponding level with respect to the frequency of the microwave field $\omega_{mw}$. 
Here the notation $\hat{\sigma}_{ij}=\ket{i}\bra{j}$ represents a projector of state $j$ onto state $i$. Note that a factor 1/2 was absorbed into the definition of the coupling strength $\Omega$ for simplicity such that $\Omega = \gamma_{e}|\mathbf{B}|/2$. Here $\gamma_{e}$ represents the electron gyromagnetic ratio ($\approx$27~GHz/T), and magnetic field  was assumed to take the form $\mathbf{B}=|\mathbf{B}|\cos(\omega_{mw}t) \hat{x}$.

The equation of motion for some arbitrary system operator $\hat{O}$ is stated in Equation \ref{eq:2_2}.
\begin{equation}
    \frac{d}{dt}\hat{O}=\frac{i}{\hbar}[\hat{H},\hat{O}]+\frac{1}{2}\sum_{n}(2\hat{L}_{n}^\dagger\hat{O}\hat{L}_{n}-\hat{O}\hat{L}_{n}^\dagger\hat{L}_{n}-\hat{L}_{n}^\dagger\hat{L}_{n}\hat{O})
    \label{eq:2_2}
\end{equation}

The operators $\hat{L}_{n}$ are jump operators representing emission, absorption and dephasing processes. Equations \ref{eq:2_1} and \ref{eq:2_2} with the model in Figure \ref{f:2_1} were used to derive a set of system equations. The jump operators used are given in Equations \ref{eq:2_3} and \ref{eq:2_4}, representing dephasing and a general transition process from level $j$ to $i$, respectively.

\begin{equation}
\hat{L}=\sqrt{\Gamma_{d}}\hat{\sigma}_{jj} \forall j \in (2,3)
\label{eq:2_3}
\end{equation}
\begin{equation}
\hat{L}=\sqrt{\gamma_{ij}}\hat{\sigma}_{ij}
\label{eq:2_4}
\end{equation}

Note the following substitutions were made for specific transitions rates (refer to Figure \ref{f:2_1}): $\gamma_{41}=\gamma_{52}=\gamma_{63}=\Lambda$, $\gamma_{14}=\gamma_{25}=\gamma_{36}=\Gamma_{sp}$, $\gamma_{98}=\Lambda_{0}$ and $\gamma_{89}=\Gamma_{sp0}$.

The set of equations derived is given as follows.
\begin{widetext}
\begin{align}
    \begin{split}
        \frac{d}{dt}\langle\hat{\sigma}_{11}\rangle = {}& i \Omega (\langle\hat{\sigma}_{31}\rangle-\langle\hat{\sigma}_{13}\rangle+\langle\hat{\sigma}_{21}\rangle-\langle\hat{\sigma}_{12}\rangle)-(\Lambda+\gamma_{21}+\gamma_{31})\langle\hat{\sigma}_{11}\rangle+\gamma_{12}\langle\hat{\sigma}_{22}\rangle\\ & +\gamma_{13}\langle\hat{\sigma}_{33}\rangle +\Gamma_{sp}\langle\hat{\sigma}_{44}\rangle+\gamma_{17}\langle\hat{\sigma}_{77}\rangle+\gamma_{19}\langle\hat{\sigma}_{99}\rangle
    \end{split}
    \label{eq:s11}
\end{align}
\begin{align}
\begin{split}
    \frac{d}{dt}\langle\hat{\sigma}_{22}\rangle = {}& i\Omega(\langle\hat{\sigma}_{12}\rangle-\langle\hat{\sigma}_{21}\rangle) +\gamma_{21}\langle\hat{\sigma}_{11}\rangle-(\Lambda+\gamma_{12})\langle\hat{\sigma}_{22}\rangle + \Gamma_{sp}\langle\hat{\sigma}_{55}\rangle+\gamma_{27}\langle\hat{\sigma}_{77}\rangle\\ & +\gamma_{29}\langle\hat{\sigma}_{99}\rangle
\end{split}    
\label{eq:s22}
\end{align}
\begin{align}
\begin{split}
    \frac{d}{dt}\langle\hat{\sigma}_{33}\rangle= {}& i\Omega(\langle\hat{\sigma}_{13}\rangle-\langle\hat{\sigma}_{31}\rangle)+\gamma_{31}\langle\hat{\sigma}_{11}\rangle-(\Lambda+\gamma_{13})\langle\hat{\sigma}_{33}\rangle+\Gamma_{sp}\langle\hat{\sigma}_{66}\rangle+\gamma_{37}\langle\hat{\sigma}_{77}\rangle\\ & +\gamma_{39}\langle\hat{\sigma}_{99}\rangle
\end{split}
\label{eq:s33}
\end{align}
\begin{equation}
    \frac{d}{dt}\langle\hat{\sigma}_{44}\rangle = \Lambda\langle\hat{\sigma}_{11}\rangle - (\Gamma_{sp}+\gamma_{74}+\gamma_{84})\langle\hat{\sigma}_{44}\rangle
    \label{eq:s44}
\end{equation}
\begin{equation}
    \frac{d}{dt}\langle\hat{\sigma}_{55}\rangle = \Lambda\langle\hat{\sigma}_{22}\rangle-(\Gamma_{sp}+\gamma_{75}+\gamma_{85})\langle\hat{\sigma}_{55}\rangle
    \label{eq:s55}
\end{equation}
\begin{equation}
    \frac{d}{dt}\langle\hat{\sigma}_{66}\rangle=\Lambda\langle\hat{\sigma}_{33}\rangle - (\Gamma_{sp}+\gamma_{76}+\gamma_{86})\langle\hat{\sigma}_{66}\rangle
    \label{eq:s66}
\end{equation}
\begin{equation}
    \frac{d}{dt}\langle\hat{\sigma}_{77}\rangle=\gamma_{74}\langle\hat{\sigma}_{44}\rangle+\gamma_{75}\langle\hat{\sigma}_{55}\rangle+\gamma_{76}\langle\hat{\sigma}_{66}\rangle-(\gamma_{17}+\gamma_{27}+\gamma_{37})\langle\hat{\sigma}_{77}\rangle
    \label{eq:s77}
\end{equation}
\begin{equation}
    \frac{d}{dt}\langle\hat{\sigma}_{88}\rangle =\gamma_{84}\langle\hat{\sigma}_{44}\rangle+\gamma_{85}\langle\hat{\sigma}_{55}\rangle+\gamma_{86}\langle\hat{\sigma}_{66}\rangle-\Lambda_{0}\langle\hat{\sigma}_{88}\rangle+\Gamma_{sp0}\langle\hat{\sigma}_{99}\rangle
    \label{eq:s88}
\end{equation}
\begin{equation}
    \frac{d}{dt}\langle\hat{\sigma}_{99}\rangle = \Lambda_{0} \langle\hat{\sigma}_{88}\rangle-(\Gamma_{sp0}+\gamma_{19}+\gamma_{29}+\gamma_{39})\langle\hat{\sigma}_{99}\rangle
    \label{eq:s99}
\end{equation}
\begin{equation}
    \frac{d}{dt}\langle\hat{\sigma}_{12}\rangle=i\Omega (\langle\hat{\sigma}_{32}\rangle+\langle\hat{\sigma}_{22}\rangle-\langle\hat{\sigma}_{11}\rangle)-\frac{1}{2}(2i\Delta_{12}+\Gamma_{d}+\gamma_{12}+\gamma_{21}+2\Lambda)
    \label{eq:s12}
\end{equation}
\begin{equation}
    \frac{d}{dt}\langle\hat{\sigma}_{13}\rangle=i\Omega (\langle\hat{\sigma}_{23}\rangle+\langle\hat{\sigma}_{33}\rangle-\langle\hat{\sigma}_{11}\rangle)-\frac{1}{2}(2i\Delta_{13}+\Gamma_{d}+\gamma_{13}+\gamma_{31}+2\Lambda)
    \label{eq:s13}
\end{equation}
\begin{equation}
    \frac{d}{dt}\langle\hat{\sigma}_{23}\rangle =i\Omega(\langle\hat{\sigma}_{13}\rangle-\langle\hat{\sigma}_{21}\rangle)+ \frac{1}{2}(2i\Delta_{12}-2i\Delta_{13}-2\Gamma_{d}-\gamma_{12}-\gamma_{13}-2\Lambda)\langle\hat{\sigma}_{23}\rangle
    \label{eq:s23}
\end{equation}
\end{widetext}
The values of the transition rates specified in the equations above were extracted from various sources. From Gupta et al. \cite{gupta_efficient_2016}, $\Gamma_{sp} = 2 \pi \times 66.16\times10^{6}$~rads/s, $\gamma_{75} = \gamma_{76} = 2\pi\times 91.8\times 10^{6}$~rads/s, $\gamma_{74} = 2\pi\times 11.1\times 10^{6}$~rads/s, $\gamma_{27} =\gamma_{37}= 2\pi\times 2.04\times 10^{6}$~rads/s, $\gamma_{17} = 2\pi\times 4.87\times 10^{6}$~rads/s. From Raman Nair et al. \cite{raman_nair_amplification_2020}, $\gamma_{19}=\gamma_{29}=\gamma_{39}=0.08\Lambda$, $\Gamma_{sp0} \approx 2\pi \times 50 \times 10^{6}$~rads/s. From Aude Craik et al. \cite{aude_craik_microwave-assisted_2020}, $\gamma_{84} = \gamma_{85} = \gamma_{86} = 0.037\Lambda$, $\Lambda_{0} = 1.3 \Lambda$. From Zhang et al. \cite{zhang_cavity_2022}, $\gamma_{s}=2\pi\times0.157$~rad/s where the longitudinal spin relaxation rates are defined as, $\gamma_{12} = \gamma_{s}(1+\bar{n})=2\pi\times 344.96$~rads/s and $\gamma_{21} = \gamma_{s}\bar{n}=2\pi\times 343.8$~rads/s. The rates were calculated at $T=300$ K which gives $\bar{n}=1/(\exp(1/(k_{B}T)\times\omega_{12}\hbar)-1)\approx 2190$. The pump rate $\Lambda$, dephasing rate $\Gamma_{d}$, MW transition frequencies $\omega_{12}$ and $\omega_{13}$, coupling strength $\Omega$ and MW driving frequency $\omega_{mw}$ can vary.
Theoretical modelling was performed to determine the effects of varying magnetic field coupling strengths on ODMR contrast. Refer to Fig. 2 in the main text for results.

\begin{figure*}
  \centering
  \includegraphics[width= 12.5cm]{"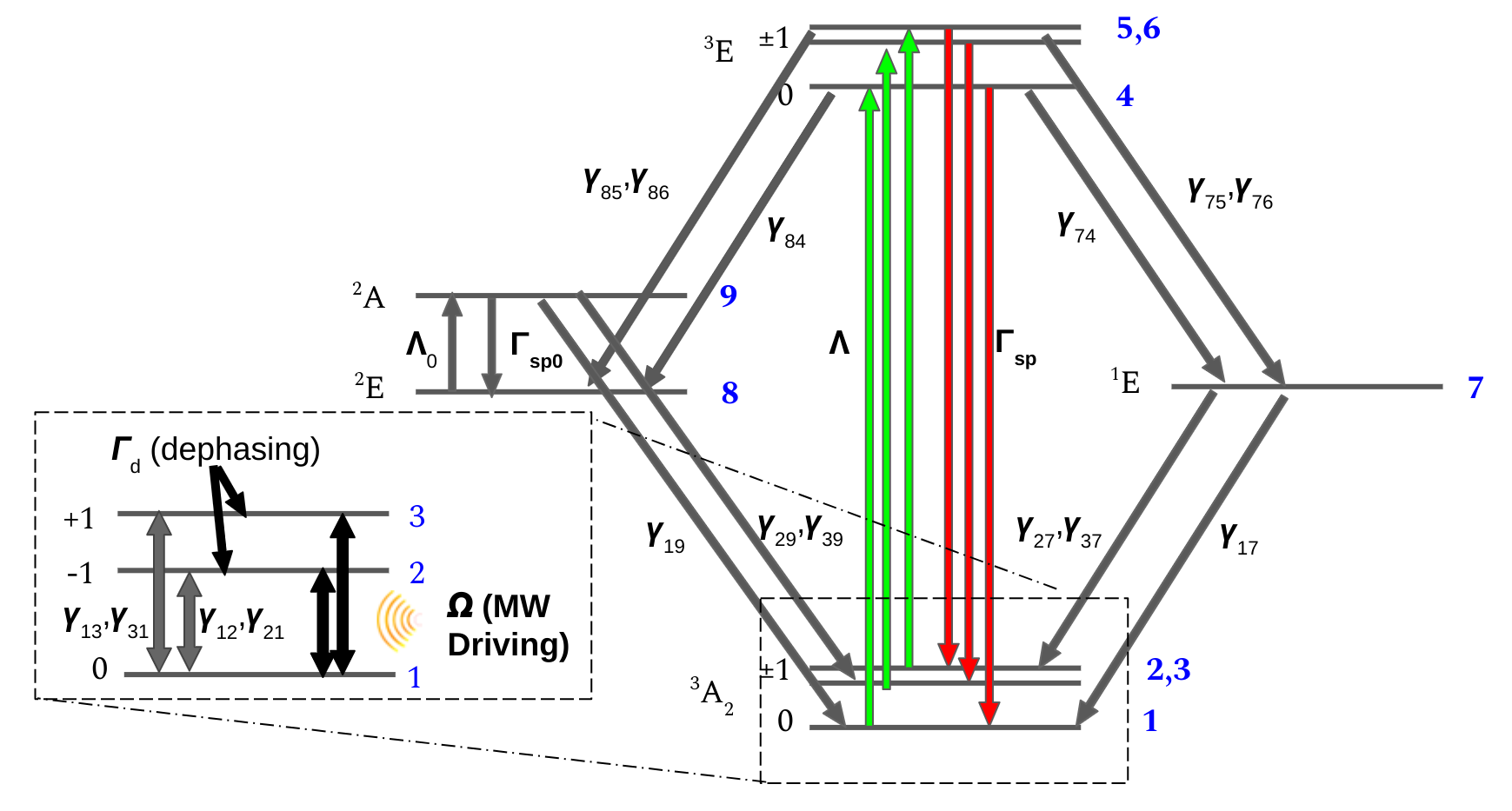"}
  \caption{Theoretical model of NV$^{-}$ spin interacting with a microwave field.} 
  \label{f:2_1}
\end{figure*}

\begin{figure*}
\includegraphics[width=12.5cm]{"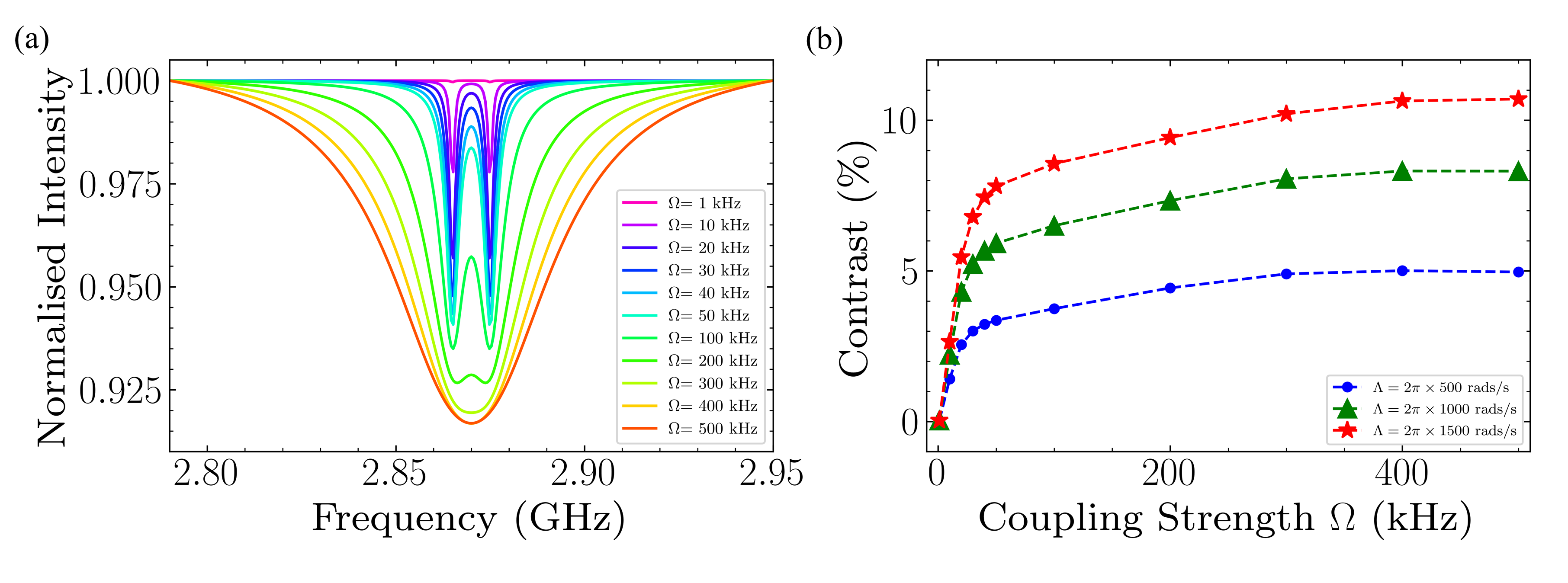"}
\caption{Numerical results from theoretical model simulations; see S.I. for further details and parameters. (a) ODMR curves simulated for varying coupling strengths ($\Omega$). (b) Contrast as a function of coupling strength plotted for three different green laser excitation powers.}
\label{f:2_2}
\end{figure*}

\section{Materials and Methods}\label{secB1}
\setcounter{figure}{0}    
\setcounter{equation}{0}
To determine the interaction between a microwave (MW) field and nitrogen-vacancy (NV) spins, we constructed a setup that to measure optically detected magnetic resonance (ODMR). A 520nm green laser (Thorlabs L520P50), operated using diode and temperature controllers, Thorlabs LDC 205 C and Thorlabs TED 200 C, respectively, delivers green light to a single-mode fibre (Thorlabs P1-460B-FC-2) using two broadband dielectric mirrors (Thorlabs BB1-E02) and a mounted aspheric lens (Thorlabs C671TME-A). The green light from this fibre was collimated using a Geltech${}^{\text{TM}}$ aspheric lens (Thorlabs C220TME-A). The laser beam is then reflected off an angled (45 degrees) 650nm long pass filter (Thorlabs FEL0650), which is then delivered into a multimode fibre (Thorlabs FG050LGA) using an achromatic doublet lens (Thorlabs AC127-025-A-ML). A fibre stripper (Micro-Electronics MS-1-FS) was used to remove the external layers from one end of the multimode fibre. A clean cut through the fibre was made using a diamond-tipped fibre-cutter (Fitel S326), which created a free and open-ended fibre core. The fibre tip was dipped into a 2~$\mu$L aqueous solution containing NV nanodiamonds (Adàmas NDNV140nmHi10ml) with 3~ppm concentration. The green laser was passed through whilst dipping, so more nanodiamonds became trapped at the beam waist due to the laser's divergence. 

The fibre tip was positioned between two white seedless grapes. The approximately ellipsoidal grapes had a major axis of $2.70\pm0.05$~cm and a minor length of $1.70\pm0.05$~cm. The grapes rested on a steel square plate with a non-conductive coating and had dimensions 60~mm x 60~mm x 3~mm thickness; see Figure 1 in the main text for an image of this setup. A 1~mm diameter copper wire 5~cm in length with enamel coating (Jaycar Electronics EN CU 1MM 18BS) was oriented vertically and passed through a 3~mm diameter hole at the centre of the plate. The copper wire was soldered on each end to the inner conductor of two stripped MW cables. An additional larger copper wire, 10~cm in length, was connected to the outer conductors of the two MW cables and passed around the plate, far away from the grape dimers. One of the MW cables was connected to an amplifier (Minicircuits ZHL-16W-43-S+), which was connected to a MW generator (Agilent Technologies E8257D). The generator outputted MW radiation with frequencies between 2.80 and 2.95~GHz. The other MW cable was connected to several attenuators, one 30~dB (Weinschel WA33-30) and two 20~dB (RS Electronics 27-9300-20), dissipating high-power loads.

The green light excited the NV spins at the fibre tip, and a fraction of red light emitted is delivered back into the fibre. The red light is filtered through the angled 650~nm long pass filter and an additional 550~nm long pass filter (Thorlabs FEL0550), which further filters any additional green light that may be present. The red light was focused using a lens (Thorlabs A220 TM-A) into a multimode fibre (Thorlabs FG050LGA) and sent to a photon detection apparatus. The red light was collimated using mounted aspheric lens (Thorlabs A280TM-B) and then aligned using two mirrors into another lens (Thorlabs AC254-35-B-ML), which focused the red light onto the 20~$\mu$m diameter photosensitive area located on the avalanche photodiode (APD) (IDQ id100-20). The APD was housed inside a box with a black cloth placed on it during operation to minimise environmental noise. The electrical signal generated by the APD was first inverted with a Picoquant SIA 400 signal inverter before being sent to a time-correlated single photon counting module (Picoquant Picoharp 300). The photon counting module was interfaced with Python, and the detected photon count rate (photons per second) was measured and recorded. With the detection of red light photons, several ODMR experiments with the green laser initiated NV spins and resonant MW field driving.

\begin{figure*}
    \centering
    \includegraphics[width=12.5cm]{"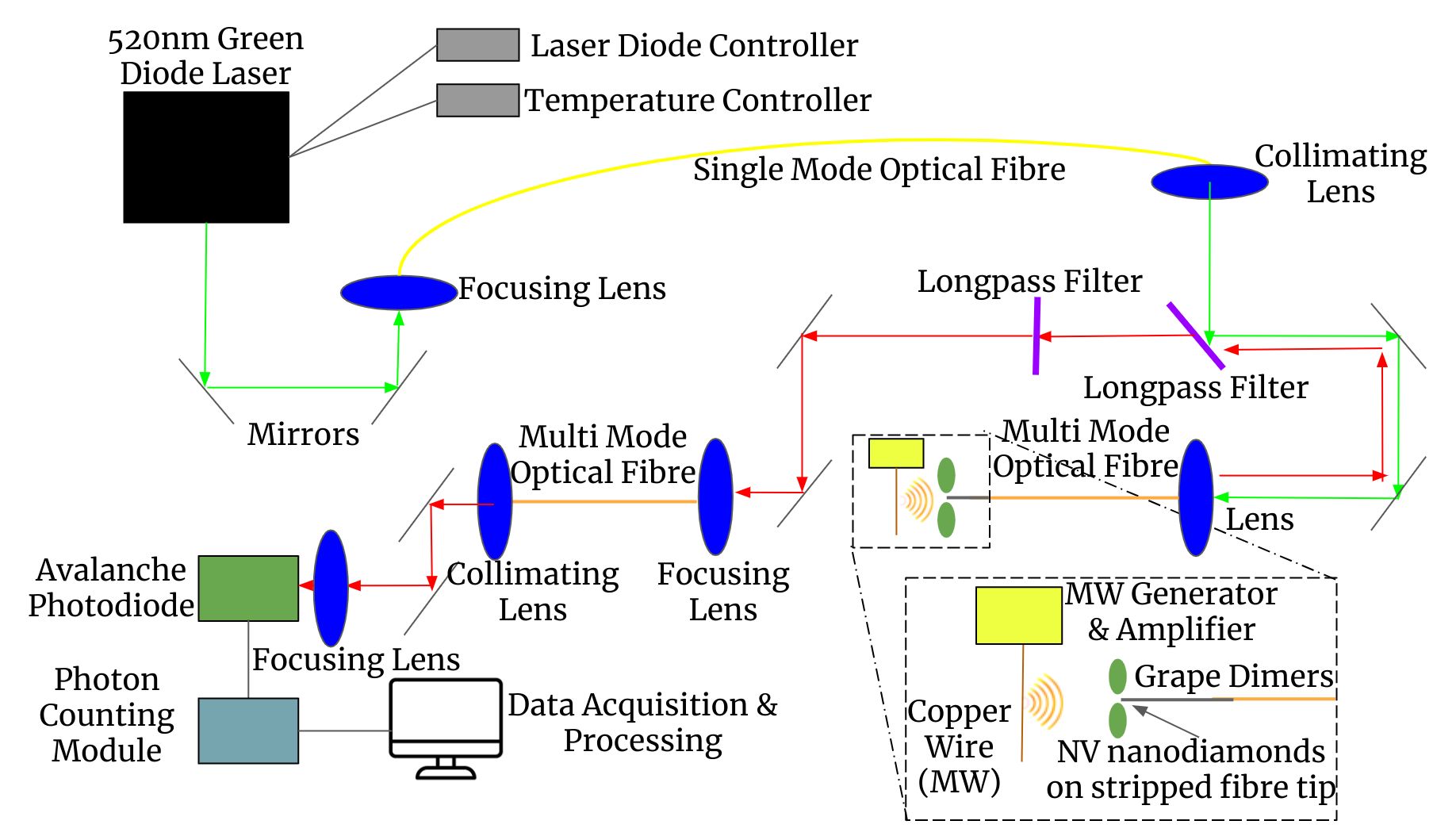"}
    \caption{Complete experimental setup to measure ODMR with NV spins using a grape dimer cavity}
    \label{fig:B1}
\end{figure*}

\section{Mie Theory Modelling}\label{secC1}

\setcounter{figure}{0}    
\setcounter{equation}{0}

Lorenz-Mie theory explains resonances based on morphology in spherical dielectric materials. The theory describes the scattering of a plane wave incident on a dielectric sphere with a diameter similar to the wavelength. Resonances typically occur when the circumference around the sphere is proportional to the wavelength \cite{hrebtov_numerical_2020}.

A plane wave propagating in the z-direction and polarised in x-direction can be described using vector spherical harmonics. The incident electric and magnetic fields are given in Equations \ref{eq:4_1} and \ref{eq:4_2}.
\begin{widetext}
\begin{equation}
\mathbf{E}_{i}=\hat{x}E_{0}e^{i(k_{2}z-i\omega t})
=E_{0}e^{-i\omega t}\sum_{n=1}^{\infty} i^{n} \frac{2n+1}{n(n+1)}(\mathbf{m}_{o1n}^{(1)}-i\mathbf{n}_{e1n}^{(1)})
\label{eq:4_1}
\end{equation}
\begin{equation}
\mathbf{H}_{i}=\hat{y}\frac{k_{2}}{\mu_{2}\omega}E_{0}e^{i(k_{2}z-i\omega t})
=-\frac{k_{2}E_{0}}{\mu_{2}\omega}e^{-i\omega t}\sum_{n=1}^{\infty} i^{n} \frac{2n+1}{n(n+1)}(\mathbf{m}_{e1n}^{(1)}+i\mathbf{n}_{o1n}^{(1)})
\label{eq:4_2}
\end{equation}
\end{widetext}
$\epsilon$ and $\mu$ are the electric permittivity and magnetic permeability, respectively. $k=\omega \sqrt{\epsilon \mu} = 2\pi/\lambda$ and $\lambda$ is the wavelength of the propagating wave. In the environment, $k_{2}=\omega\sqrt{\epsilon_{2}\mu_{2}}=\omega n_{2}/c$ with refractive index $n_{2}$. Similarly, in the sphere is $k_{1}=\omega\sqrt{\epsilon_{1}\mu_{1}}=\omega n_{1}/c$ where $n_{1}$ is the refractive index of the sphere. Furthermore, $E_{0}$ is the amplitude, and the vector spherical harmonics that appear in the electric and magnetic field expressions are given in Equations \ref{eq:4_3} and \ref{eq:4_4}.

\begin{equation}
\mathbf{m}_{emn/omn}=\nabla\times(\mathbf{r}\psi_{emn/omn})
\label{eq:4_3}
\end{equation}

\begin{equation}
\mathbf{n}_{emn/omn}=\frac{\nabla\times\mathbf{m}_{emn/omn}}{k}
\label{eq:4_4}
\end{equation}

where $\psi_{emn}=cos(m\phi) P^{m}_{n}(\cos{\theta}) z_{n}(kr)$ and $\psi_{omn}=sin(m\phi) P^{m}_{n}(\cos{\theta}) z_{n}(kr)$. Here $P^{m}_{n}(\cos{\theta})$ are the associated Legendre polynomials, and $z_{n}(kr)$ are spherical Bessel functions or spherical Hankel functions where specified. $\theta$ and $\phi$ are the polar and azimuthal angles in spherical coordinates, respectively. 

The fields internal to the sphere (r$<$R) are:

\begin{equation}
\mathbf{E}_{int}=E_{0}e^{-i\omega t}\sum_{n=1}^{\infty} i^{n} \frac{2n+1}{n(n+1)}(a^{int}_{n}\mathbf{m}_{o1n}^{(1)}-i b^{int}_{n}\mathbf{n}_{e1n}^{(1)})
\label{eq:4_5}
\end{equation}
\begin{equation}
\mathbf{H}_{int}=-\frac{k_{1}E_{0}}{\mu_{1}\omega}e^{-i\omega t}\sum_{n=1}^{\infty} i^{n} \frac{2n+1}{n(n+1)}(b^{int}_{n}\mathbf{m}_{e1n}^{(1)}+i a^{int}_{n}\mathbf{n}_{o1n}^{(1)})
\label{eq:4_6}
\end{equation}

The fields external to the sphere (r$>$R) are:
\begin{equation}
\mathbf{E}_{ext}=E_{0}e^{-i\omega t}\sum_{n=1}^{\infty} i^{n} \frac{2n+1}{n(n+1)}(a^{ext}_{n}\mathbf{m}_{o1n}^{(2)}-i b^{ext}_{n}\mathbf{n}_{e1n}^{(2)})
\label{eq:4_7}
\end{equation}
\begin{equation}
\mathbf{H}_{ext}=-\frac{k_{2}E_{0}}{\mu_{2}\omega}e^{-i\omega t}\sum_{n=1}^{\infty} i^{n} \frac{2n+1}{n(n+1)}(b^{ext}_{n}\mathbf{m}_{e1n}^{(2)}+i a^{ext}_{n}\mathbf{n}_{o1n}^{(2)})
\label{eq:4_8}
\end{equation}

The superscripts (1) and (2) specifies the function $z_n(kr)$ as the spherical Bessel function of the first kind or the spherical Hankel function, respectively. 

The coefficients in the expressions above are given as follows.

\begin{equation}
a^{int}_{n}=\frac{\mu_{1}[\rho h_{n}(\rho)]'j_{n}(\rho)-\mu_{1}[\rho j_{n}(\rho)]'h_{n}(\rho)}{\mu_{1}[\rho h_{n}(\rho)]' j_{n}(N\rho)-\mu_{2}[N\rho j_{n}(N\rho)]' h_{n}(\rho)}
\label{eq:4_9}
\end{equation}
\begin{equation}
b^{int}_{n}=\frac{\mu_{1}N[\rho h_{n}(\rho)]'j_{n}(\rho)-\mu_{1}N[\rho j_{n}(\rho)]'h_{n}(\rho)}{\mu_{2}N^{2}[\rho h_{n}(\rho)]' j_{n}(N\rho)-\mu_{1}[N\rho j_{n}(N\rho)]'h_{n}(\rho)}
\label{eq:4_10}
\end{equation}
\begin{equation}
a^{ext}_{n}=-\frac{\mu_{1} j_{n}(N\rho)[\rho j_{n}(\rho)]'-\mu_{2}j_{n}(\rho)[N\rho j_{n}(N\rho)]'}{\mu_{1} j_{n}(N\rho)[\rho h_{n}(\rho)]'-\mu_{2} h_{n}(\rho)[N\rho j_{n}(N\rho)]'}
\label{eq:4_11}
\end{equation}
\begin{equation}
b^{ext}_{n}=-\frac{\mu_{1} j_{n}(\rho)[N\rho j_{n}(N\rho)]'-\mu_{2}N^{2}j_{n}(N\rho)[\rho j_{n}(\rho)]'}{\mu_{1}h_{n}(\rho)[N\rho j_{n}(N\rho)]'-\mu_{2} N^{2}j_{n}(N\rho)[\rho h_{n}(\rho)]'}
\label{eq:4_12}
\end{equation}

Magnetic field resonance requires minimising the denominator of $a^{int/ext}_{n}$. This gives the characteristic equation in Equation \ref{eq:4_13}

\begin{equation}
\mu_{1} j_{n}(N\rho)[\rho h_{n}(\rho)]'-\mu_{2} h_{n}(\rho)[N\rho j_{n}(N\rho)]'\approx 0
\label{eq:4_13}
\end{equation}

\begin{figure}
    \centering
    \begin{subfigure}{0.5\textwidth}
    \includegraphics[width=5cm]{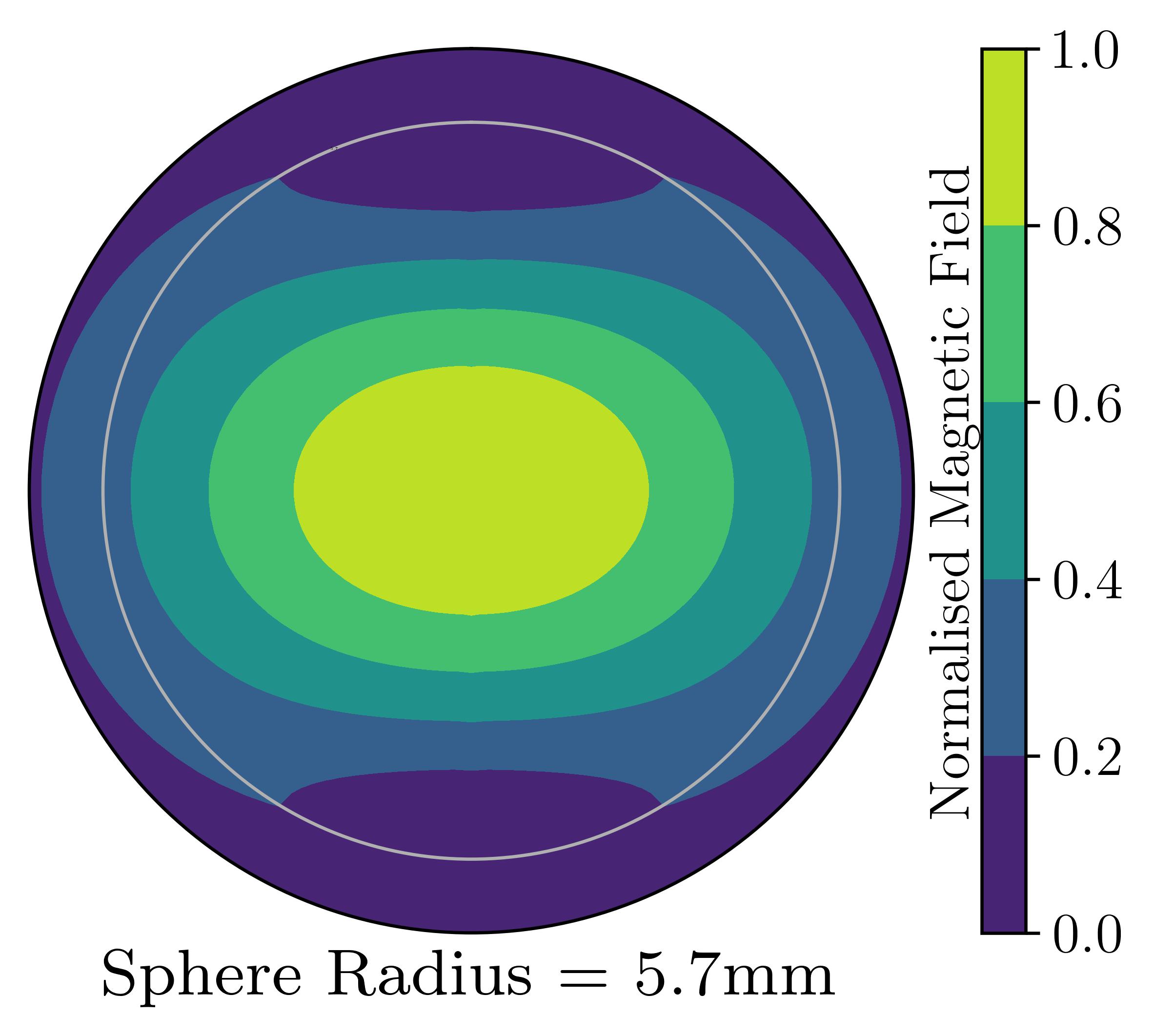}
    \caption{}
    \label{fig:4_1a}
    \end{subfigure}
    \begin{subfigure}{0.5\textwidth}
    \includegraphics[width=5cm]{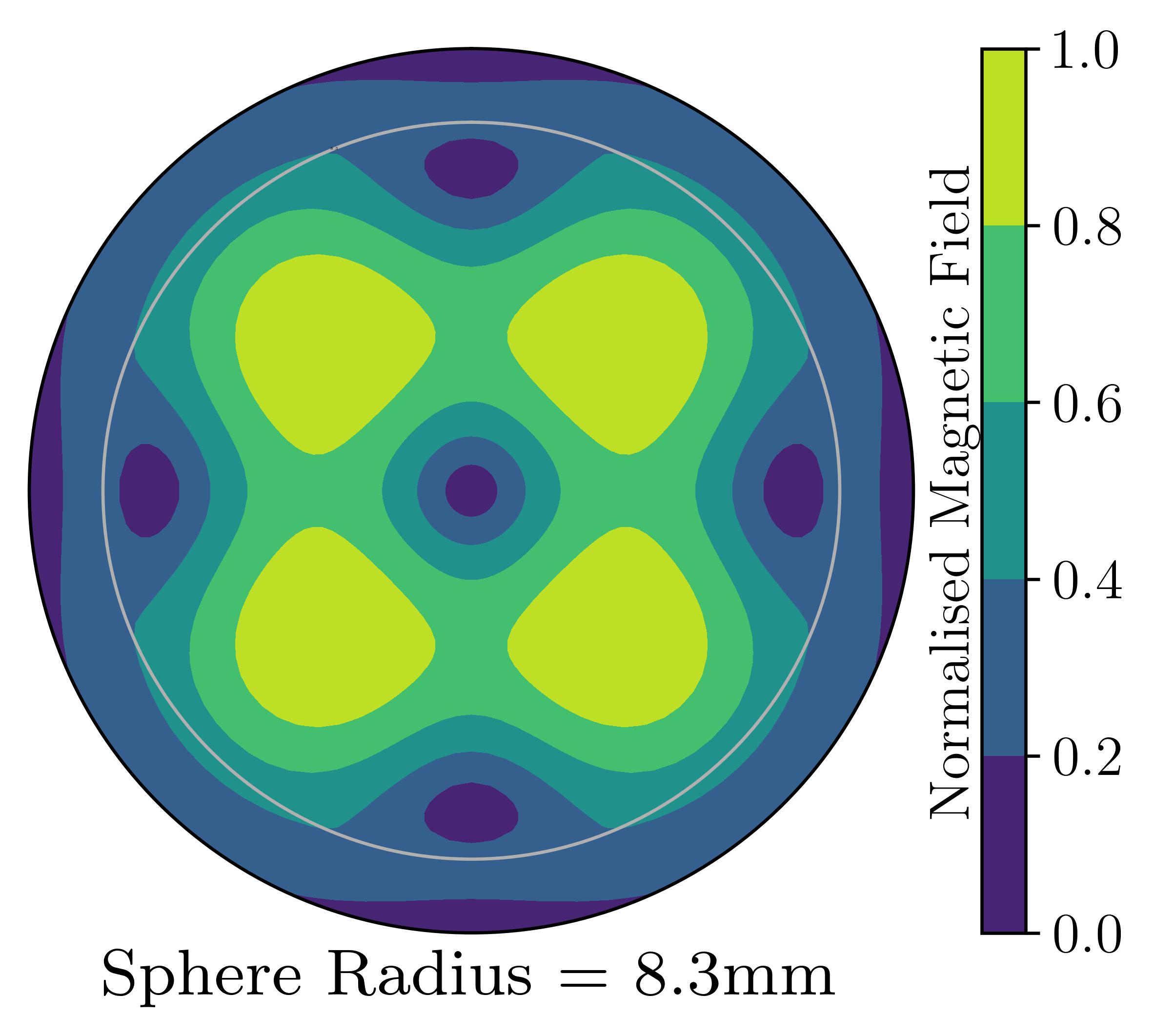}
    \caption{}
    \label{fig:4_1b}
    \end{subfigure}
    \begin{subfigure}{0.5\textwidth}
    \includegraphics[width=5cm]{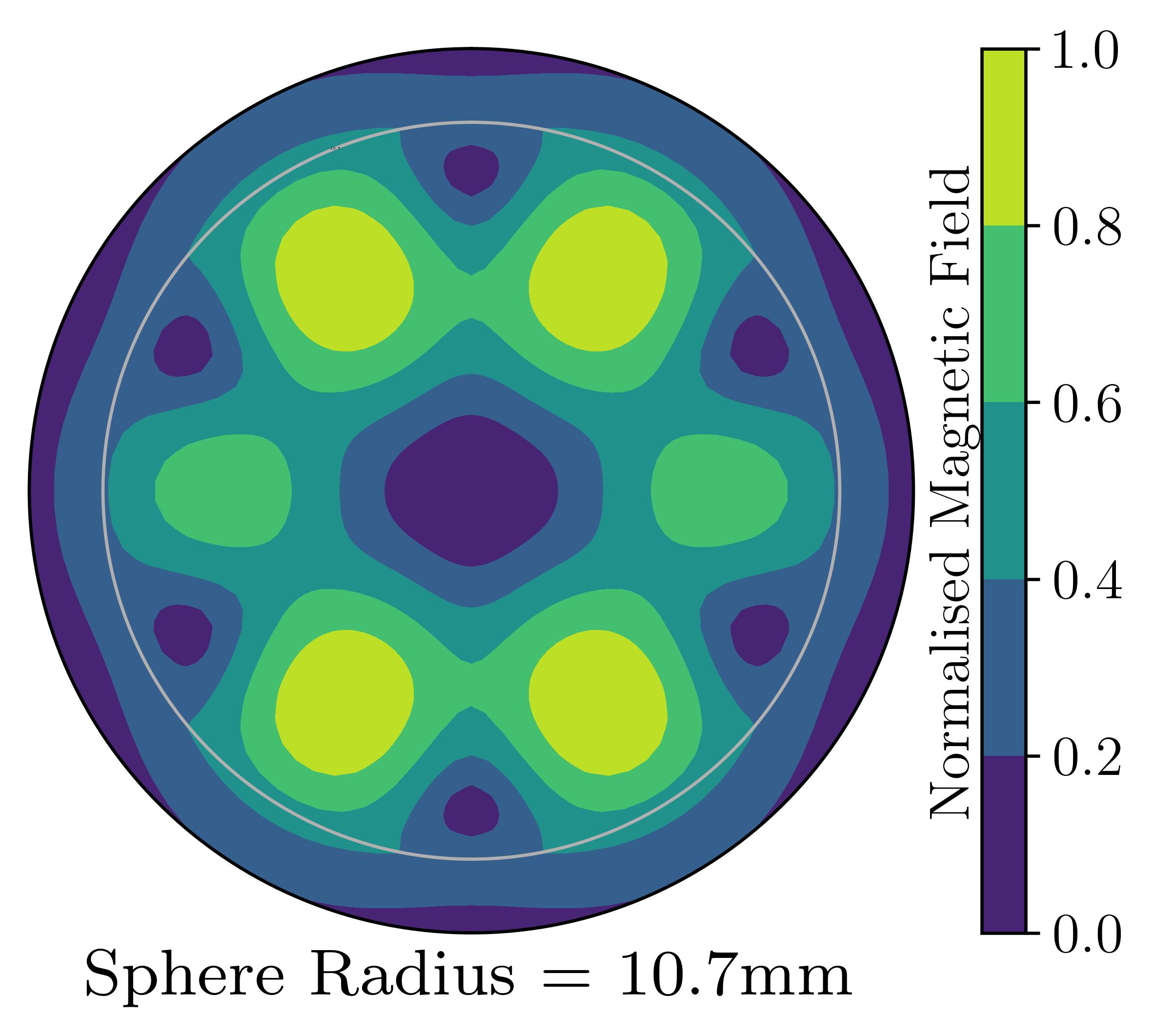}
    \caption{}
    \label{fig:4_1c}
    \end{subfigure}
    \caption{Mie theory analysis of spherical magnetic field resonances at 2.87 GHz. The radius of each sphere is shown in each figure. (a) 1st Mode (n=1), (b) 2nd Mode (n=2), (c) 3rd Mode (n=3).}
    \label{fig:4_1}
\end{figure}

Assuming a refractive index of water at MW frequency $\approx2.9$~GHz is $n_{1}=8.9$ \cite{khattak_linking_2019} and $n_{2}=1$ for the environment, using Equation \ref{eq:4_13}, the radius of the spheres for which resonance first occurs at roughly 2.87~GHz (NV$^{-}$ ground state zero-field splitting) for the first three modes (n=1,2,3). The magnetic field at a cross-section is plotted for the first three modes in Figure \ref{fig:4_1}.

Given a solution ($\rho_{res}$) to Equation \ref{eq:4_13}, the circumference can be related to the wavelength as $C = 2\pi R= \rho_{res}\lambda$. Grapes are typically non-spherical and more akin to ellipsoids. The size of the ellipsoids chosen for the ODMR experiments was based on the range of grape sizes available and the relationship between the perimeter and resonant wavelength. For an ellipsoid, the resonance condition in terms of the perimeter of an ellipse (Ramanujan approximation) is given by Equation \ref{eq:4_14}. 

\begin{equation}
    C_{elps} \approx \pi(a+b)\left(1+\frac{3h}{10+\sqrt{4-3h}}\right)=\alpha\lambda
    \label{eq:4_14}
\end{equation}
 
where $h=(a-b)^{2}/(a+b)^{2}$, $\alpha$ is some constant of proportionality, $a$ and $b$ are the semi-major axis and semi-minor axis length of the ellipse, respectively. As an assumption, $\alpha$ was assumed to be equivalent to $\rho$ for the case of the sphere at resonance. This assumption is equivalent to equating the perimeter around a sphere to the perimeter around an ellipsoid (in the major axis direction), given that they are composed of the same material. The size of the grapes chosen for experiments was based on the $n=3$ mode (i.e. $\alpha=0.645$), where the appropriate values of a and b were determined given available grape sizes.
\vfill
\bibliography{references}% Produces the bibliography via BibTeX.

\end{document}